# Potential of Atmospheric Pressure Thermal Plasma Technology towards Waste Processing: A Comprehensive review

*Tejashwi Rana, Aishik Basu Mallick, Radhika T P, Suryasunil Rath, Pratyay Chattopadhyay, Satyananda Kar**

*Plasma Applications Laboratory, Department of Energy Science and Engineering, Indian Institute of Technology Delhi, Hauz Khas, New Delhi, India-110016*

## Abstract

The enhancement of living standards has significantly contributed to the rapid growth of urban populations, resulting in a substantial increase in municipal solid waste (MSW) generation. This trend underscores the critical need for sustainable, environmentally friendly, cost-effective, and highly efficient waste management solutions. This study highlights the pressing necessity for effective MSW management and examines plasma pyrolysis/gasification as an emerging technology to address this challenge. The article provides a detailed analysis of thermal plasma generation techniques employing diverse power sources, including direct current, alternating current, radiofrequency inductively coupled, and microwave-based systems. A comparative evaluation of various plasma torch designs is conducted, emphasizing their applicability in waste-to-energy and waste treatment processes. A comprehensive overview of the treatment of a broad spectrum of waste materials, such as MSW, sewage sludge, coal, wood, plastics, tyres, and rubber, using thermal arc plasma technology is presented. The process predominantly converts waste into a combustible gas (syngas) with a calorific value ranging from 5 to 15 MJ/Nm³ and produces vitrified slag or ash as a by-product. The findings suggest that thermal plasma pyrolysis/gasification offers a promising approach to waste management, facilitating energy generation and material recovery while addressing the challenges of increasing MSW generation.

## 1 Introduction

Millions of tons of municipal solid waste (MSW) are generated daily, and its management has become a significant concern. It deserves special attention because of its negative environmental, health, and economic externalities [1]. Currently, the annual generation of MSW is around 2 billion tons worldwide [2]. The projected data shows an increment of 30% (2.6 billion tons) by 2030 & 70% (3.4 billion tons) by 2050, from the current figure considering the constant growth rate in GDP and population. Countries responsible for most of the total waste generation are India, China, and the United States, which generate around 40% . The global waste treatment and disposal report estimated that most generated waste was either landfilled without gas collection or openly dumped. In 2016, a report estimated that solid waste management had produced 1.6 billion tonnes of $CO_2$-equivalent greenhouse gas emissions, accounting for around 5% of world emissions. The emissions may rise to 2.6 billion tonnes of $CO_2$-equivalent by 2050 unless the sector improves [3]. Currently, a widely accepted technology used for MSW disposal is incineration. However, the demand for excess airflow limits the required temperature in incineration. The low temperature in the process chamber generates toxic pollutants like Chlorinated dioxins and furans, leading to air pollution and dreaded diseases like cancer and abnormalities [4]. New technologies like pyrolysis, conventional gasification, and plasma pyrolysis/gasification are in practice to minimize the adverse effects on the environment [5]. Plasma pyrolysis is an emerging and eco-friendly technology for waste management, converting waste to valuable products without affecting the environment. It may be defined as the decomposition/treatment of waste using plasma as an energy input source by gasification in an oxygen-starved environment [6]. It offers various advantages over other conventional technologies regarding environmental safety and may be considered a promising waste management technology. It provides a high level of destruction efficiency and is eco-friendly [4]. It simultaneously gasifies the combustible waste fractions into syngas for energy generation and non-combustible fractions into vitrified slags. It also has the advantages of reducing emissions to zero and converting waste into a valuable product or by-product [7]. This article reviews the significant aspects of different sources for plasma generation, plasma pyrolysis/gasification technology, and the benefits of plasma-assisted treatment in the waste management sector. The current review focuses on providing a better understanding of this novel eco-friendly technology.

## 2 Plasma

Plasma is generally considered the fourth state of matter after solid, liquid, and gas. More than 99% of the known universe is in the plasma state. The plasma state can be natural or artificial. Natural plasmas are the sun, lightning strikes, aurora borealis, etc. Some examples of artificial plasmas are fluorescent tubes and television screens. Plasma consists of electrons, ions, and neutral particles, and the existence of high-energy particles relative to ordinary gases facilitates numerous applications. Neutrals and ions are heavy particles compared to electrons, and they may present in the excited state because of the high energy content of plasma. These exited particles return to the ground state and emit photons responsible for plasma luminosity. The plasma is in an equilibrium state between negative and positive charges (electrically neutral). This property of plasma is known as quasi-neutrality [8].

## 2.1 Thermal plasma generation

Plasmas are generated by providing energy to neutral gases, which produce charge carriers like electrons and ions due to collision between high-energy electrons/photons and atoms/molecules of neutrals. The energy can be provided in the form of mechanical energy through adiabatic compression of gas, chemical energy through an exothermic chemical reaction, or electrical fields that can be applied or through beams of electrons/ions/neutrals. The usual method for plasma generation is applying an external electric field to neutral gas for electrical breakdown. Every space includes cosmic particles in its background and has charge carriers. These charge carriers are accelerated under the applied field and collide with the atoms/molecules of neutrals. This collision generates an avalanche of charged particles, eventually balancing the loss of charge carriers and generating steady-state plasma [9].

Classifying plasma into thermal and non-thermal is based on electric discharges. A different characteristic of both the plasmas is given in Table 1. Thermal plasma, also known as hot plasma or equilibrium plasma, is high energy density plasma produced generally by Direct Current (DC) or Radio Frequency (RF) inductively coupled plasma (ICP). Non-thermal plasma, known as cold or non-equilibrium plasma, is lower energy density plasma produced in the glow, low-pressure RF, and corona discharges [10–12].

**Table 1. Characteristics of Thermal and Non-thermal Plasma**

| | | | |
|---|---|---|---|
| Thermal/equilibrium/hot plasma | High energy density | $T_e = T_h$ | Electron density – $10^{23}$ – $10^{28}$ $m^{-3}$<br>Electron temperature – 1-2 eV |
| Non-thermal/non-equilibrium/cold | Lower energy density | $T_e \geq T_h$ | Electron density < $10^{20}$ $m^{-3}$<br>Electron temperature – few eV |

$T_e$ – electron temperature, $T_h$ – Heavy particle temperature (ions & neutrals)

Thermal plasma generation occurs through various discharges like DC/AC discharge, RF inductively coupled discharge, and Microwave (MW) discharge. Table 2 shows a comparative study of different thermal plasma. A DC arc discharge generates a high energy density and a high-temperature zone between two electrodes with a high plasma gas flow rate [13]. The arc may be transferred or of a non-transferred type. In a transferred arc, plasma is confined in between the electrodes, while in a non-transferred arc, the plasma generation occurs between two electrodes and comes outside for different applications. In the case of arc discharges, the torch and electrodes are generally water-cooled, and life is around 200-500 hours. The available power of the torch is 1.5 MW, and scaling is possible up to 6 MW. On the other hand, RF plasma torch can be inductive or capacitive; transferring energy to plasma gas in the form of electromagnetic can provide high energy input per unit volume. Radio frequency and microwaves can transmit energy via resonant energy transfer from the waves to the plasma. The torch comparatively has more life since it is not exposed to severe plasma conditions. The power levels available for the RF torch are 100 kW, and scaling is possible up to 1 MW [14–16].

**Table 2: Comparison between different types of thermal plasma**

| Parameters | DC arc plasma | RF ICP plasma | Microwave plasma |
|---|---|---|---|
| Temperature | 5000-10000 K | 3000-8000 K | 1200-2000 K |
| Electrode erosion | Yes | No | No |
| Cooling of torch | Required for tungsten,<br>Not required for graphite | Required | Not required |
| Plasma ignition | Easy | Difficult | Difficult |
| Plasma volume | Small | Medium | Large |
| Plasma velocity | High | Low | Low |
| Power supply efficiency (%) | 60-90 | 40-70 | 40-70 |
| Plasma torch power | 1.5 MW | 100 KW | - |

## 2.2 Fundamental heat transfer mechanism in thermal plasma

The thermal plasma's Local Thermal Equilibrium property makes it considered a fluid with specific thermodynamic and transport properties. Therefore, the phenomenon of conduction, convection, and radiation can explain the heat transfer mechanism. Thermal plasma has higher enthalpy, thermal conductivity, and radiation intensity than other fluids; thus, heat transfer is extremely high. The overall heat transfer mechanism includes heat energy transfer from plasma to waste material through conduction, convection and radiation, and from waste material to the surroundings (towards the reactor wall) through the radiation. The net heat energy required to decompose the waste material is the difference between the heat transfer (convection and radiation are dominant) to the material from plasma and the loss of radiative heat energy from the material to the surroundings given in equation 1.

$$Q_{net} = hA(T_p - T_s) + Aq_r - \sigma\varepsilon A\,(T_s^4 - T_a^4) \quad (1)$$

Where h is the heat transfer coefficient of the material, A is the surface area of the material, $q_r$ is radiative heat transfer from plasma to material, and $T_p$, $T_s$, and $T_a$ are the temperature associated with plasma, material surface, and reactor wall, respectively, $\sigma$ is the Stephan-Boltzmann constant, $\varepsilon$ is the emissivity of the material [17]. The waste material observes the heat energy from plasma and vaporizes, forming a gas and resulting in a heat transfer mechanism [14].

## 3 Thermal plasma for waste treatment

Thermal plasmas are applied to treat a wide range of wastes, from municipal solid waste (MSW) to sewage sludge waste (SSW), harmful organic compounds, e-waste and medical wastes, etc [18–20]. They transfer high heat fluxes of high temperatures and reactive species, which allows the destruction of any type of waste [14]. Moreover, thermal plasmas provide distinctive advantages such as high enthalpy, high reactivity, oxidation and reduction atmosphere, and rapid quenching rate, which lead to [21]:

a. High heat transfer rate
b. melting of high-temperature materials
c. Low gas flow rate
d. smaller installation size for a given waste throughput
e. The feasibility of producing saleable byproducts

The uniqueness and advantages of thermal plasma-assisted waste treatment always push further exploration, as shown in Figure 1. Plasma waste treatment has been a hot topic for the last decade. The reason behind this might be the issues generated by the volume of waste generation, environmental carbon footprint, and global warming from waste dumping in landfills/open dumping demands for sustainable and eco-friendly technology.

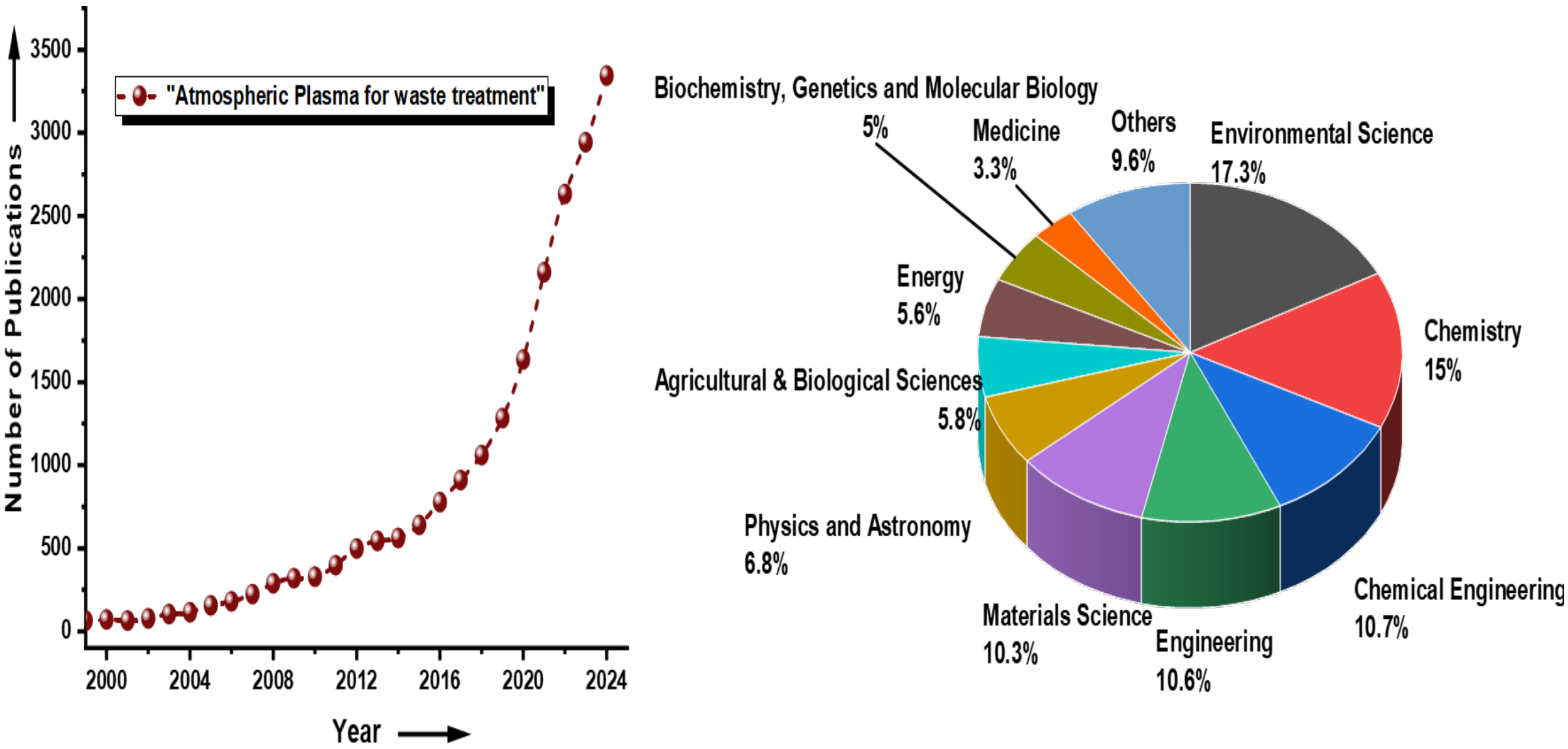

Fig 1. Left: Journal articles published over the years, Right: Journal articles published in different subject areas with the keyword "Atmospheric Plasma for waste treatment" as per the statistical data available in the SCOPUS (abstract and citation database) [22] database.

Thermal Plasma classification is based on different power sources such as DC, AC, RF-ICP, and MW. Detailed drawings of various types of thermal arc plasma sources for waste treatment application are given in Figure 2. All these sources have been used at small to industrial scales in previous years to treat various types of waste and convert it into a valuable product and by-products. Different sources have their advantages and disadvantages but in comparison to the conventional methods, these technologies have proven themselves as a better solution in waste management. One significant barrier to commercializing these technologies is the capital cost of higher energy consumption. The detailed discussion will be discussed in upcoming sections.

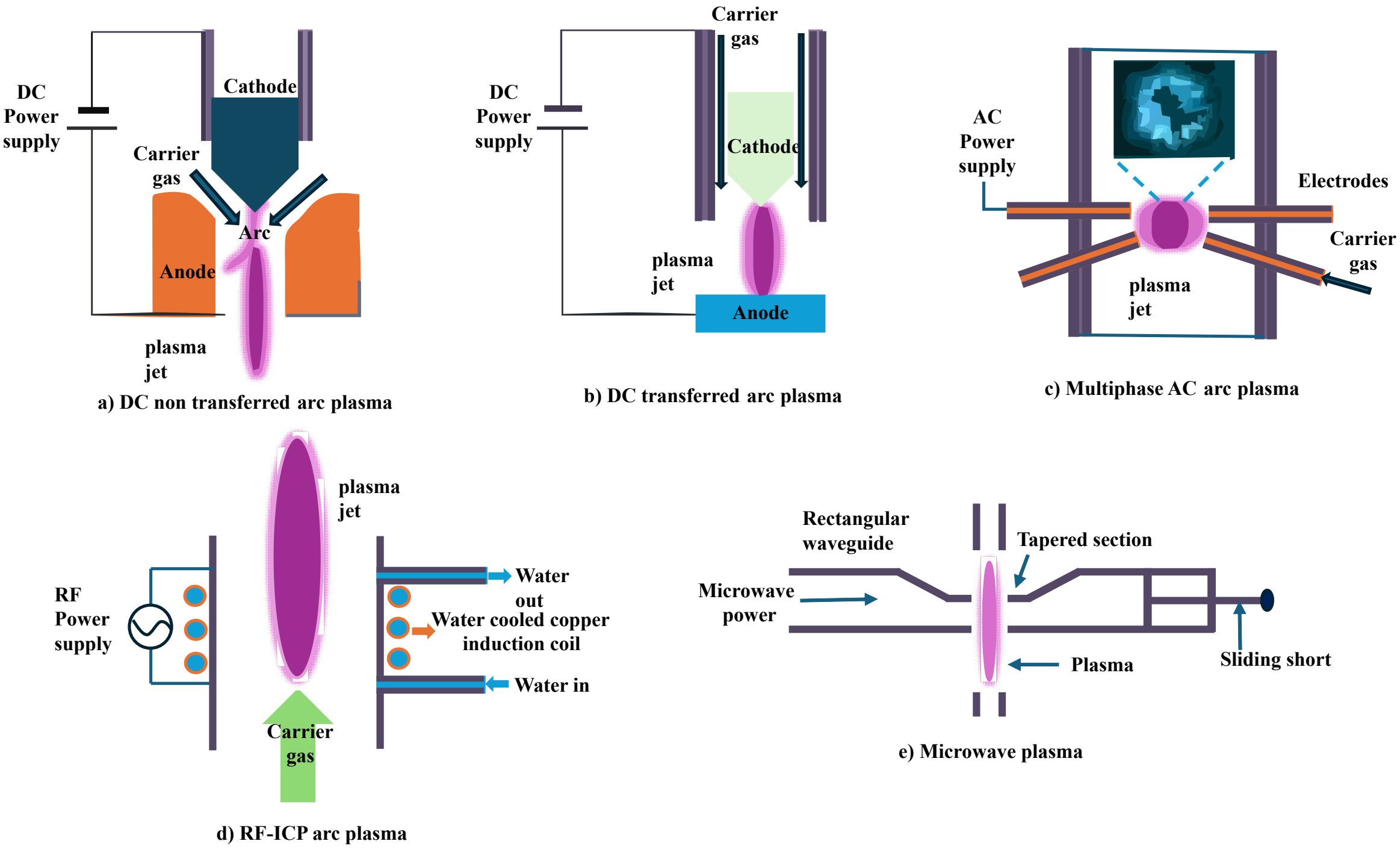

Fig 2. Different types of thermal plasma sources: a) DC non-transferred arc plasma, b) DC transferred arc plasma, c) Multiple AC arc plasma, d) RF-ICP arc plasma, and e) Microwave plasma for waste treatment application [23–25]

### 3.1 DC Thermal Arc Plasma

Generally, the thermal plasmas are generated by an electric arc for waste treatment because of its nonreactive nature towards the processing environment [14]. These thermal arcs are produced using two electrodes, carrier gas, and applied potential difference. The applied potential difference breaks down the gas between the electrode gap and forms an arc, generating thermal plasma. The electrodes are comprised of highly electrically conductive materials such as copper, tungsten, graphite, etc. The most common carrier gases are nitrogen, argon, air, helium, oxygen, steam, etc [26]. Air as a plasma carrier gas is commonly used for waste treatment to supply sufficient oxygen for proceeding reactions (gasification), cost optimization, and reduced nitrogen content (syngas) in output gases. However, oxygen as a carrier gas has a limitation of a higher electrode erosion rate. In some cases, $N_2$, $CO_2$, and steam are also supplied to the torch for plasma generation at the cost of high arc voltage and high electrode erosion rate particularly in the case of steam because of the generation of a mixture of $H_2$, $O_2$, and OH radicals [27]. The life of the electrode is the major concern in DC arc plasma as it eroded due to the high-temperature zone with the processing time. The life can be minimised using argon as carrier gas but lower specific heat and thermal conductivity than others lead to low power output and reduce the application in waste treatment. The thermal plasma arc is mainly of two types transferred & non-transferred, and the temperature of the arc is in the range of 10000-20000 K. The temperature of the arc depends on input power, plasma carrier gas, and plasma torch configurations [28].

Different types of DC plasma torches, such as Westinghouse, Europlasma, Tetronics, and PSC, are available for waste treatment, with a capacity of up to 100 MW [29]. These thermal plasma torches treated a wide range of waste and converted it into valuable products and by-products, details of which are mentioned in Table 3. Hrabovsky *et al*. [30] conducted a detailed investigation of waste such as plastic, coal, fire sawdust, wooden pellets, RDF, pyrolysis oil, etc. The plasma power varies between 100 and 140 kW in a chamber with 0.2 $m^3$ volume and wall temperature between 1100 – 1400 °C. Hydrogen is found at maximum (61%) in the case of coal with a steam flow rate of 18g/min, and a lower heating value of produced gas is found at 11.3 $MJ/m^3$. Yousef *et al.* [31] use a DC steam plasma torch for plasma gasification of surgical masks (10 kg/run) for hydrogen-rich syngas generation. The hydrogen produced approximately 50% with other combustible gases such as CO (22%) and methane (13%) at a power input of 56.9 kW, where syngas generated with a rate of 3.4 $m^3$/kg and lower heating value is 14.5 $MJ/m^3$. Ibrahim *et al.* [32] treated municipal solid waste through plasma gasification technology in a temperature range of 600 – 1600 °C in a 10 kg/h plasma chamber with an argon (carrier gas) flow rate of 10 LPM. The treatment conducted in the presence of oxygen and steam is more prone to CO generation, where maximum generation is 95% in a temperature range of 1200 to 1400 °C. Nema *et al.* [33] and Rana *et al.* [34] use graphite electrodes for DC arc plasma waste treatment for municipal solid waste treatment and convert them into combustible gases and non-hazardous ash. Graphite electrodes have some advantages over conventional plasma torches, such as water cooling is not needed as graphite can resist higher temperature up to 3000 °C leads to lower input energy requirement than conventional plasma torch. Mallick *et al.* [35–37] use graphite electrode-based $CO_2$ plasma gasification of various wastes such as bakelite electrical switches, ABS plastic, and RDF for the generation of high-value syngas (13.95 MJ/m$^3$) and liquid oil (30.59 MJ/kg).

**Table 3: Different types of waste treated through DC thermal arc plasma technology**

| Waste treated | Carrier gas | Power input and feed | output | Reference |
|---|---|---|---|---|
| RDF | $CO_2$ | 1.6 kW, 40 gm/run | $H_2$ (43 %), CO (44 %) | [35] |
| ABS Plastic | $CO_2$ | 1.6 kW, 40 gm/run | $H_2$ (30 %), CO (46 %) | [36] |
| Bakelite | $CO_2$ | 1.6 kW, 40 gm/run | $H_2$ (24 %), CO (42 %) | [37] |
| Polypropylene | $N_2$ | 35.2 kW, 3.6 kg/h | $H_2$(54 %), CO (3 %), $CH_4$(6 %) | [38] |
| MSW | Air | 30 kW, 50 kg/h | $H_2$ (8 %), CO (21 %) | [34] |
| Tyre | $N_2$ | 35.2 kW, 4.8 kg/h | $H_2$ (50 %), CO (11 %) | [39] |
| Rubber | $N_2$ | 35.2 kW, 2.6 kg/h | $H_2$ (45 %), CO (16 %) | [40] |
| Medical waste | $N_2$ | 50 kW, 25 kg/h | $H_2$ (41 %), CO (48 %) | [33] |
| Surgical mask | steam | 56.9 kW, 10 kg/h | $H_2$(50 %), CO (22 %), $CH_4$(13 %) | [31] |
| petrochemical | Ar | 3.6 kW, 10 cm$^3$ | $H_2$(33 %), CO (16 %), $CH_4$(17 %) | [41] |
| Carpet | Ar | 100 kW, 256 gm | $H_2$ (13 %), CO (20 %) | [42] |

### 3.2 AC thermal arc plasma

In this category, a combination of electrode systems uses different AC frequency (50-60 Hz). The torch consists of cylindrical geometry with electrodes at an oblique angle to the gas flow. The plasma gas passes through the annular space between the electrodes, where electric power input is dissipated via the Joule effect, generating a high-temperature plasma plume. This energy is transferred to the main flow through convection, diffusion, and radiation. The resulting plasma plume typically exhibits extremely high velocities (several hundred meters per second), driven by the combination of the narrow nozzle design and the intense thermal expansion of the heated plasma gas. In this system, multiple arcs coexist within a large discharge chamber, creating a less confined plasma than classical DC plasma torches. This configuration is sustained by a continuous, high concentration of charge carriers due to the relatively long lifetime of ionized species despite alternating current and zero-current crossings. Consequently, the plasma achieves a lower mean temperature and operates in a diffuse mode to some extent, unlike the contracted mode in DC systems. Arc motion, driven by alternating currents and magnetohydrodynamic (MHD) forces, enhances convection transfer. Energy is distributed through gas heating, relaxation of excitation, and atom recombination, reducing radiation and convection losses. These features result in higher electro-thermal efficiency compared to DC plasma torches.

Several types of waste are treated using AC thermal plasma and converted into a valuable product and by-product, given in detail in Table 4. Rutberg *et al.* [43] treated plastic waste and converted it into syngas using a 50 Hz three-phase steam plasma torch for plasma gasification. The plasma arc diameter and discharge length are 4.47 mm and 498 mm, respectively. The thermal arc temperature ranges from 10,000-11,500 K with a torch efficiency of 94.3–95.3%. The steam and air are supplied in a 1:6 ratio, and syngas are generated at a flow rate of 3.48–3.62 m³/kg, where voltage varies in a range of 1145–1853 V. Institute for Electrophysics and Electric Power Russian Academy of Sciences (IEE RAS) group [44] has developed different torch configurations with mixtures of gases from air, steam, and methane. Surov *et al.* [45] treated coal, wood, and RDF using air plasma gasification. The characteristics obtained a plasma arc temperature of 8300-10500 K, and arc dimension of 600 mm long, and 6-8 mm in diameter.

**Table 4 : Different Wastes treated through AC plasma**

| Waste treated | Carrier gas | Power input and feed | Output | References |
|---|---|---|---|---|
| Plastic waste | Steam, air | 57.65–87.54 kW | $H_2$ (62 %), CO (34 %) | [43] |
| Coal | Steam, $CO_2$, $CH_4$ | 5-150 kW | $H_2$ (22.96 %), CO (20.46 %) | [45] |
| Wood | Steam, $CO_2$, $CH_4$ | 5-150 kW | $H_2$ (28.20 %), CO(26.87 %) | [45] |
| RDF | Steam, $CO_2$, $CH_4$ | 5-150 kW | $H_2$ (16.95 %), CO(27.30 %) | [45] |

### 3.3 RF-ICP thermal plasma

RF-ICP systems generate plasma using a high-frequency current, typically in the range of 2 to 27 MHz, flowing through a coil surrounding the plasma chamber [46]. Induces an electric field within the chamber, which initiates a ring-shaped current inside the reactor, producing plasma. The generated plasma achieves high energy densities ($10^5$–$10^7$ W/cm$^2$) and operating temperatures ranging from 1000 K to 10,000 K. Such conditions, combined with high ionized particle concentrations ($10^{16}$–$10^{17}$ particles/cm$^3$), foster rapid reaction rates supported by the presence of free ions and excited molecules. Increases heat and mass transfer, making RF-ICP suitable for waste pyrolysis and gasification [18,47]. A key advantage of RF-ICP systems is their ability to utilize a wide variety of carrier gases, air, $O_2$, $H_2$, and even hydrocarbons, such as methane, along with Ar, Ar/$H_2$, Ar/He, and Ar/$N_2$. Argon is a commonly used gas in RF-ICP Torches due to its low heat capacity at ionizing temperatures, making it highly efficient for initiating plasma formation [8]. Its relatively high density also contributes to effective thermal conductivity, and its affordability makes it a practical choice. However, maintaining gas purity is crucial, as impurities can quench the plasma torch, causing instability or failure in plasma generation. To overcome limitations posed by argon's density, lighter gases such as hydrogen, helium, or nitrogen are often introduced into the system [48,49]. These gases, being less dense, enhance the axial propulsion of the plasma jet, improving performance and enabling a more focused plasma output. This flexibility, combined with the scalability of RF ICP systems to power levels > 1 MW, allows them to meet the demands of high-throughput commercial operations. Furthermore, their continuous operation is unaffected by issues such as current-carrying limitations or cooling requirements associated with DC systems [50].

However, RF-ICP systems have certain limitations. The energy transfer efficiency of induction heating is inherently lower than that of DC systems, primarily due to the lower efficiency of high-frequency power supplies. Traditional RF-ICP systems often rely on vacuum tube-driven Class C-tuned oscillators, which have less than 50% efficiency. Nevertheless, advancements in solid-state power supplies, operating at frequencies up to ~450 kHz, are improving the efficiency of high-power RF-ICP systems. Additionally, careful optimization of waste injection locations and gas flow rates is essential to maintain the performance and stability of the plasma [28]. Hybrid plasma systems have been developed to overcome the limitations of individual plasma configurations by combining their strengths. Radio Frequency (RF) plasma systems are valued for their high purity, large plasma volumes, and low gas velocity, making them ideal for precise applications like material processing and waste treatment. However, RF systems face challenges such as difficult ignition and susceptibility to extinction under certain conditions. In contrast, Direct Current (DC) plasma systems offer stable operation and easy ignition, ensuring reliability across various scenarios. By integrating the stability and ignition ease of DC systems and control of RF systems, hybrid DC-RF plasma systems can provide a versatile solution [51].

Various types of waste, such as heavy oil, tyre powder, and municipal solid waste (MSW) combined with raw wood, are being effectively processed using Radio Frequency Inductively Coupled Plasma (RF-ICP) technology, with nitrogen ($N_2$) commonly serving as the carrier gas. The input power for these treatments varies significantly, ranging from 0.3 kW to 10 kW, allowing adaptability across different waste materials. For example, research conducted by Tang *et al*. [52] investigated the pyrolysis of waste tyres in an RF-ICP reactor, operating at power levels between 1600 and 2000 W and maintaining reactor pressures of 3000 to 8000 Pa. This setup achieved a reactive plasma environment with temperatures between 1200 K and 1800 K and employed dual-electrode systems to improve the yield of gases like hydrogen ($H_2$) and carbon monoxide (CO). Similarly, Aboughaly *et al*. [53] introduced a novel RF thermal plasma pyrolysis apparatus with a remarkable 89 wt.% conversion efficiency without producing tar. Their system featured an 8-turn copper RF-ICP torch within a 12 L thermochemical reactor, capable of maintaining stable operation for over 30 minutes at peak temperatures reaching 2000 °C. The pyrolysis process was conducted at temperatures ranging from 550 °C to 990 °C for 30 minutes, while gasification occurred at 1300 °C for 1 second. Additionally, Shie *et al*. [54] explored the gasification of MSW mixed with raw wood, reporting syngas yields between 88.59% and 91.84 vol%, with energy recovery rates ranging from 59.07% to 111.89%, underscoring the effectiveness of RF-ICP technology in advancing sustainable waste management practices. Table 5 summarizes the scalability of RF-ICP for different waste types.

**Table 5 : Different types of waste treated through RF ICP technology**

| Waste treated | Carrier gas | Power input and feed | output | Reference |
|---|---|---|---|---|
| Tyre | $N_2$ | 1.8 kW, 0.012 kg/h | $H_2$ (47 %), CO (17 %) | [52] |
| Rice Straw | $N_2$ | 0.3-0.7 kW | $H_2$ (2.2 mg/L), CO (25.9 mg/L), | [55] |
| Tire Powder | $N_2$ | 1.6-2 kW, 0.012 k g/h | $H_2$, CO, $CH_4$, $CO_2$; Pyrolytic Char | [56] |
| MSW and wood | $N_2$ | 10 kW, 0.2 kg/hr | CO and $H_2$; 88.59 - 91.84 vol% | [54] |

### 3.4 Microwave thermal plasma

Microwave plasmas, generated by high-frequency electromagnetic waves (300 MHz-10 GHz), are commonly employed in waste treatment applications. MW heating offers several advantages over conventional methods, such as more uniform heat distribution and improved control over the heating process [57]. Various microwave-coupled plasma torches are widely used and capable of functioning across a broad power range, from a few watts to several hundred kilowatts, and under diverse gas pressures, from low to atmospheric pressure. MW plasmas exhibit significantly higher electron densities compared to other low-frequency plasmas. This high electron density leads to the extensive dissociation of the working gas, making it highly chemically reactive. Additionally, most MW plasma does not require electrodes, eliminating contamination risks. This electrode-free design ensures a longer service life and reduced maintenance requirements. The different types of MW atmospheric pressure plasma jets (MW-APPJs), reactor design, and plasma parameters are discussed further [25].

Uhm *et al*. [58] used a swirl-type gasifier equipped with two MW steam plasmas that achieved an inner-wall temperature of up to 1700°C, gasifying Indonesian brown coal with high ash content (33.17%). At a chamber temperature of 1640°C, the system demonstrated a nearly 100% carbon conversion rate and an impressive cold gas efficiency (ratio of syngas heat content to fuel heat content in ambient conditions) of 84%, which is exceptionally high for a relatively small gasifier. Lin *et al*. [59] developed an environmentally friendly method to recycle aluminium dross to produce high-value alumina products. The process successfully dissolved 46% of secondary aluminium dross (SAD) with NaOH to form $NaAl(OH)_4$ solution, while 54% remained as residues. High-purity aluminium trihydrate precipitate was obtained by adjusting the solution to pH 7. Using atmospheric pressure MW plasma calcination, they produced high-purity α-$Al_2O_3$ (98% pure), representing about 25% of the original aluminium mass in SAD. Sturm *et al*. [60] conducted a study that demonstrated that MW -driven plasma gasification could successfully convert biomass waste into fuel gas, though residence time was too short for complete conversion. The process generated a surplus in energy, with the fuel gas output containing up to 1.84 times more energy than the input MW energy, representing an 84% energy surplus. While plasma stability posed challenges under certain conditions, the results suggest this system could enable effective small-scale waste treatment when combined with fuel cells for energy recovery. The different waste treatment using MW plasma are summarized in Table 6.

**Table 6 : Different Wastes treated through Microwave plasma**

| Waste | Carrier gas | Power input and feed | Output | References |
|---|---|---|---|---|
| Polyethylene | Ar | 0.8 kW, 5 g/run | $H_2$ (9.5%), CO (72.6%) | [61] |
| RDF | Ar | 0.8 kW, 5 g/run | $H_2$ (13.8%), CO (65.5%) | [61] |
| Glycerol | $N_2$ | 1-2 kW, 180 g/h | $H_2$ (57%), CO (35%) | [62] |
| Spirulina algae | $N_2$ | 0.8-1 kW | $H_2$ (22.8%), CO (4.11%), $CO_2$ (31.6%) | [63] |
| Sewage sludge | $CO_2$ | 1.4 kW, 10 g/run | $H_2$ (31.74–35.28%), CO (14.22–25.26%) | [64] |
| Coal | Air | 10 kW, 100 kg/h | $H_2$ (33.6%), CO (26.6%) | [65] |
| Biomass | Air/$N_2$ | 6 kW | $H_2$ (41%), CO (53%) | [66] |
| Brown Coal | Steam | 4 kW | $H_2$ (48%), CO (23%) | [67] |

## 4 Conclusion

Thermal plasma pyrolysis represents a cutting-edge approach in modern waste treatment technologies, offering significant energy and material recovery potential. This study highlights the versatility of thermal plasma torches for waste-to-energy applications. Laboratory-scale to industrial-scale investigations demonstrate the feasibility of using various plasma torches (DC/AC/RF-ICP/MW) for treating diverse waste streams. At the same time, DC plasma torches are recommended for industrial-scale operations for higher efficiency and operational reliability. However, in DC torches, major capital and operating costs go to power electronics switching for rectification of AC-DC power [68]. Also, for a general topology of tip-to-cylindrical geometries, it has been observed that the electrodes used in DC torches undergo erosion due to several mechanical and thermal stresses, reducing their operational lifetime. On the contrary, multiphase AC torches promise the possibility of large-scale industrial applications such as waste treatment, pyrolysis and gasification by using high-current transformers. Compared to conventional DC plasma arc generators, RF ICP systems also present unique advantages and operational flexibilities [17,26]. While DC plasma systems are favoured in hazardous waste treatment due to their stable operation and reduced refractory wear, they require frequent maintenance due to electrode erosion. Additionally, reactive gases (oxygen) are often incompatible with specific electrode materials, such as graphite in DC systems, limiting their operational scope. In contrast, RF-ICP and MW plasma torches are electrodeless, eliminating the challenges associated with electrode erosion, contamination by metallic vapours, and routine maintenance. Despite substantial advancements in recent years, developing and optimizing thermal plasma pyrolysis and gasification processes for large-scale applications face notable technical and economic challenges. The increasing emphasis on environmental protection and the conservation of energy and resources is expected to accelerate the advancement and industrial adoption of this technology in the near future.

**Acknowledgements**

The authors acknowledge Invest India and Principal Scientific Advisor (PSA) to the Government of India (GoI) for financial assistance in conducting the research under the Waste to Wealth mission. The author also acknowledges the plasma group of the Department of Energy Science and Engineering (DESE), IIT Delhi for their support and guidance.

**References**

[1] A.D. Diaz-Barriga-Fernandez, J.E. Santibañez-Aguilar, N. Radwan, F. Nápoles-Rivera, M.M. El-Halwagi, J.M. Ponce-Ortega, Strategic Planning for Managing Municipal Solid Wastes with Consideration of Multiple Stakeholders, ACS Sustain. Chem. Eng. 5 (2017) 10744–10762. https://doi.org/10.1021/acssuschemeng.7b02717.
[2] World bank, Trends in solid waste management, 2021. https://datatopics.worldbank.org/what-a-waste/trends_in_solid_waste_management.html.
[3] S. Kaza, L.C. Yao, P. Bhada-Tata, F. Van Woerden, What a Waste 2.0: A Global Snapshot of Solid Waste Management to 2050, 2018. https://doi.org/10.1596/978-1-4648-1329-0.
[4] Central Pollution Control Board, Study on Plastic Waste Disposal through “Plasma Pyrolysis Technology,” 110032 (2016) 1–53. https://cpcb.nic.in/displaypdf.php?id=cGxhc3RpY3dhc3RlL1BsYXNtYS1QeXJvbHlzaXMtZmluYWwtUmVwb3J0LTIxLTExLTE2LnBkZg==.
[5] J. Oakey, Fuel flexible energy generation: Solid, liquid and gaseous fuels, Woodhead Publishing, 2015.
[6] M.T. Munir, I. Mardon, S. Al-Zuhair, A. Shawabkeh, N.U. Saqib, Plasma gasification of municipal solid waste for waste-to-value processing, Renew. Sustain. Energy Rev. 116 (2019) 109461. https://doi.org/10.1016/j.rser.2019.109461.
[7] J. Li, K. Liu, S. Yan, Y. Li, D. Han, Application of thermal plasma technology for the treatment of solid wastes in China: An overview, Waste Manag. 58 (2016) 260–269. https://doi.org/10.1016/j.wasman.2016.06.011.
[8] M.I. Boulos, P. Fauchais, E. Pfender, Thermal Plasmas Fundamentals and Applications volume 1, 1994.
[9] H. Conrads, M. Schmidt, Plasma generation and plasma sources, Plasma Sources Sci. Technol. 9 (2000) 441–454. https://doi.org/10.1088/0963-0252/9/4/301.
[10] R. T P, S. Kar, Glow-to-arc discharge transitions in a radio frequency atmospheric pressure plasma jet, Phys. Fluids 36 (2024). https://doi.org/10.1063/5.0218872.
[11] R. T P, S. Kar, Effect of an additional floating electrode on radio frequency cross-field atmospheric pressure plasma jet, Sci. Rep. 13 (2023) 10665. https://doi.org/https://doi.org/10.1038/s41598-023-37805-7.
[12] A.B. Mallick, G.V. Prakash, S. Kar, R. Narayanan, Development of a pulse modulated sub-radio frequency power supply for atmospheric pressure plasma devices, Rev. Sci. Instrum. 94 (2023). https://doi.org/10.1063/5.0173873.
[13] A. Yadav, S. Karmakar, S. Kar, M. Kumar, Numerical modelling of a direct current non-transferred thermal plasma torch for optimal performance, Contrib. to Plasma Phys. 63 (2023) e202200088. https://doi.org/10.1002/ctpp.202200088.
[14] V.S. Sikarwar, M. Hrabovský, G. Van Oost, M. Pohořelý, M. Jeremiáš, Progress in waste utilization via thermal plasma, Prog. Energy Combust. Sci. 81 (2020). https://doi.org/10.1016/j.pecs.2020.100873.
[15] L. Tang, H. Huang, H. Hao, K. Zhao, Development of plasma pyrolysis/gasification systems for energy efficient and environmentally sound waste disposal, J. Electrostat. 71 (2013) 839–847. https://doi.org/10.1016/j.elstat.2013.06.007.
[16] H. Huang, L. Tang, Treatment of organic waste using thermal plasma pyrolysis technology, Energy Convers. Manag. 48 (2007) 1331–1337. https://doi.org/10.1016/j.enconman.2006.08.013.
[17] E. Gomez, D.A. Rani, C.R. Cheeseman, D. Deegan, M. Wise, A.R. Boccaccini, Thermal plasma technology for the treatment of wastes: A critical review, J. Hazard. Mater. 161 (2009) 614–626. https://doi.org/10.1016/j.jhazmat.2008.04.017.
[18] B. Ruj, S. Ghosh, Technological aspects for thermal plasma treatment of municipal solid waste - A review, Fuel Process. Technol. 126 (2014) 298–308. https://doi.org/10.1016/j.fuproc.2014.05.011.
[19] R.C. Sanito, S.-J. You, Y.-F. Wang, Application of plasma technology for treating e-waste: A review, J. Environ. Manage. 288 (2021) 112380. https://doi.org/https://doi.org/10.1016/j.jenvman.2021.112380.
[20] X. Cai, C. Du, Thermal Plasma Treatment of Medical Waste, Plasma Chem. Plasma Process. 41 (2021) 1–46. https://doi.org/10.1007/s11090-020-10119-6.
[21] S. Safa, G. Soucy, Liquid and solution treatment by thermal plasma: A review, Int. J. Environ. Sci. Technol. 11 (2014) 1165–1188. https://doi.org/10.1007/s13762-013-0356-3.
[22] Elsevier, SCOPUS, (n.d.). https://www.scopus.com/results/results.uri?st1=atmospheric+plasma+for+waste+treatment&st2=&s=ALL%28atmospheric+AND+plasma+AND+for+AND+waste+AND+treatment%29&limit=10&origin=searchbasic&sort=plf-f&src=s&sot=b&sdt=b&sessionSearchId=9a666baf3b00231e2d7c7b2d2.
[23] K.I.M. Soonho, Decomposition of Pharmaceuticals and Personal Care Products by Water Plasma, Kyushu University Institutional Repository, 2022. https://hdl.handle.net/2324/5068200.
[24] E. Gomez, D.A. Rani, C.R. Cheeseman, D. Deegan, M. Wise, A.R. Boccaccini, Thermal plasma technology for the

treatment of wastes: A critical review, J. Hazard. Mater. 161 (2009) 614–626. https://doi.org/https://doi.org/10.1016/j.jhazmat.2008.04.017.
[25] S. Rath, S. Kar, Microwave atmospheric pressure plasma jet: A review, Contrib. to Plasma Phys. n/a (n.d.) e202400036. https://doi.org/https://doi.org/10.1002/ctpp.202400036.
[26] H.A. Gabbar, S.A. Darda, V. Damideh, I. Hassen, M. Aboughaly, D. Lisi, Comparative study of atmospheric pressure DC, RF, and microwave thermal plasma torches for waste to energy applications, Sustain. Energy Technol. Assessments 47 (2021) 101447. https://doi.org/10.1016/j.seta.2021.101447.
[27] M. Hrabovsk\`y, Generation of thermal plasmas in liquid-stabilized and hybrid dc-arc torches, Pure Appl. Chem. 74 (2002) 429–433.
[28] J. Heberlein, A.B. Murphy, Thermal plasma waste treatment, J. Phys. D. Appl. Phys. 41 (2008) 53001. https://doi.org/10.1088/0022-3727/41/5/053001.
[29] F. Fabry, C. Rehmet, V. Rohani, L. Fulcheri, Waste gasification by thermal plasma: a review, Waste and Biomass Valorization 4 (2013) 421–439. https://doi.org/10.1007/s11090-015-9616-y.
[30] M. Hrabovsky, M. Hlina, M. Konrad, V. Kopecky, T. Kavka, O. Chumak, A. Maslani, Thermal plasma gasification of biomass for fuel gas production, High Temp. Mater. Process. An Int. Q. High-Technology Plasma Process. 13 (2009). https://doi.org/10.1615/HighTempMatProc.v13.i3-4.40.
[31] S. Yousef, A. Tamošiūnas, M. Aikas, R. Uscila, D. Gimžauskaitė, K. Zakarauskas, Plasma steam gasification of surgical mask waste for hydrogen-rich syngas production, Int. J. Hydrogen Energy 49 (2024) 1375–1386. https://doi.org/10.1016/j.ijhydene.2023.09.288.
[32] İ. Yayalık, A. Koyun, M. Akgün, Gasification of Municipal Solid Wastes in Plasma Arc Medium, Plasma Chem. Plasma Process. 40 (2020) 1401–1416. https://doi.org/10.1007/s11090-020-10105-y.
[33] S.K. Nema, K.S. Ganeshprasad, Plasma pyrolysis of medical waste, Curr. Sci. 83 (2002) 271–278. https://doi.org/jstor.org/stable/24106885.
[34] T. Rana, S. Kar, Assessment of energy consumption and environmental safety measures in a plasma pyrolysis plant for eco-friendly waste treatment, J. Energy Inst. 114 (2024) 101617. https://doi.org/https://doi.org/10.1016/j.joei.2024.101617.
[35] R. Mallick, P. Vairakannu, Experimental studies on CO2-thermal plasma gasification of refused derived fuel feedstock for clean syngas production, Energy 288 (2024) 129766. https://doi.org/10.1016/j.energy.2023.129766.
[36] R. Mallick, P. Vairakannu, Experimental investigation of acrylonitrile butadiene styrene plastics plasma gasification, J. Environ. Manage. 345 (2023) 118655. https://doi.org/10.1016/j.jenvman.2023.118655.
[37] R. Mallick, P. Vairakannu, CO2 plasma gasification of bakelite-based electrical switch waste feedstock, J. Clean. Prod. 423 (2023) 138813. https://doi.org/10.1016/j.jclepro.2023.138813.
[38] L. Tang, H. Huang, Z. Zhao, C.Z. Wu, Y. Chen, Pyrolysis of polypropylene in a nitrogen plasma reactor, Ind. \& Eng. Chem. Res. 42 (2003) 1145–1150.
[39] L. Tang, H. Huang, Thermal plasma pyrolysis of used tires for carbon black recovery, J. Mater. Sci. 40 (2005) 3817–3819.
[40] H. Huang, L. Tang, C.Z. Wu, Characterization of gaseous and solid product from thermal plasma pyrolysis of waste rubber, Environ. Sci. \& Technol. 37 (2003) 4463–4467. https://doi.org/https://doi.org/10.1021/es034193c.
[41] H. Karimi, M.R. Khani, M. Gharibi, H. Mahdikia, B. Shokri, Plasma pyrolysis feasibility study of spent petrochemical catalyst wastes to hydrogen production, J. Mater. Cycles Waste Manag. 22 (2020) 2059–2070. https://doi.org/doi.org/10.1007/s10163-020-01089-0.
[42] A. Vaidyanathan, J. Mulholland, J. Ryu, M.S. Smith, L.J. Circeo Jr, Characterization of fuel gas products from the treatment of solid waste streams with a plasma arc torch, J. Environ. Manage. 82 (2007) 77–82. https://doi.org/10.1016/j.jenvman.2005.12.006.
[43] P.G. Rutberg, V.A. Kuznetsov, E.O. Serba, S.D. Popov, A. V Surov, G. V Nakonechny, A. V Nikonov, Novel three-phase steam--air plasma torch for gasification of high-caloric waste, Appl. Energy 108 (2013) 505–514. https://doi.org/10.1016/j.apenergy.2013.03.052.
[44] A.N. Bratsev, V.E. Popov, A.F. Rutberg, S. V Shtengel', A facility for plasma gasification of waste of various types, High Temp. 44 (2006) 823–828. https://doi.org/10.1007/s10740-006-0099-7.
[45] A. V. Surov, S.D. Popov, V.E. Popov, D.I. Subbotin, E.O. Serba, V.A. Spodobin, G. V. Nakonechny, A. V. Pavlov, Multi-gas AC plasma torches for gasification of organic substances, Fuel 203 (2017) 1007–1014. https://doi.org/10.1016/j.fuel.2017.02.104.
[46] M. Boulos, J. Mostaghimi, Thermal Plasma Sources: How Well are They Adopted to Process Needs?, Plasma Chem. \& Plasma Process. 35 (2015). https://doi.org/10.1007/s11090-015-9616-y.
[47] V. Colombo, A. Concetti, E. Ghedini, M. Gherardi, P. Sanibondi, B. Vazquez, Rf Thermal Plasma Vitrification of Incinerator Bottom and Fly Ashes with Waste Glasses from Fluorescent Lamps, Ispc_20 (2011) 2–5.
[48] D.M. McClenathan, W.C. Wetzel, S.E. Lorge, G.M. Hieftje, Effect of the plasma operating frequency on the figures of merit of an inductively coupled plasma time-of-flight mass spectrometer, J. Anal. At. Spectrom. 21 (2006) 160–167. https://doi.org/10.1039/B515719F.
[49] T.B. Reed, Induction - Coupled Plasma Torch , 824 (2008) 821–824. https://doi.org/https://doi.org/10.1063/1.1736112.
[50] A. You, M. Be, I. In, High temperature–high pressure thermal conductivity of argon, 3947 (2020) 3939–3947. https://doi.org/https://doi.org/10.1063/1.437946.
[51] K. Kawajiri, T. Sato, H. Nishiyama, Experimental analysis of a DC–RF hybrid plasma flow, Surf. Coatings Technol. 171 (2003) 134–139. https://doi.org/https://doi.org/10.1016/S0257-8972(03)00256-1.
[52] L. Tang, H. Huang, Some observations from studies of RF plasma pyrolysis of waste tires, Chem. Eng. Commun. 197 (2010) 1541–1552. https://doi.org/10.1080/00986445.2010.485013.

[53] M. Aboughaly, H.A. Gabbar, V. Damideh, I. Hassen, RF-ICP Thermal Plasma for Thermoplastic Waste Pyrolysis Process with High Conversion Yield and Tar Elimination, Processes 8 (2020). https://doi.org/10.3390/pr8030281.
[54] J.-L. Shie, L.-X. Chen, K.-L. Lin, C.-Y. Chang, Plasmatron gasification of biomass lignocellulosic waste materials derived from municipal solid waste, Energy 66 (2014) 82–89. https://doi.org/https://doi.org/10.1016/j.energy.2013.12.042.
[55] W.-K. Tu, J.-L. Shie, C.-Y. Chang, C.-F. Chang, C.-F. Lin, S.-Y. Yang, J.T. Kuo, D.-G. Shaw, Y.-D. You, D.-J. Lee, Products and bioenergy from the pyrolysis of rice straw via radio frequency plasma and its kinetics, Bioresour. Technol. 100 (2009) 2052–2061. https://doi.org/https://doi.org/10.1016/j.biortech.2008.09.052.
[56] H. Huang, L. Tang, Pyrolysis treatment of waste tire powder in a capacitively coupled RF plasma reactor, Energy Convers. Manag. 50 (2009) 611–617. https://doi.org/https://doi.org/10.1016/j.enconman.2008.10.023.
[57] E. Tatarova, F.M. Dias, E. Felizardo, J. Henriques, M.J. Pinheiro, C.M. Ferreira, B. Gordiets, Microwave air plasma source at atmospheric pressure: Experiment and theory, J. Appl. Phys. 108 (2010). https://doi.org/10.1063/1.3525245.
[58] H.S. Uhm, Y.H. Na, Y.C. Hong, D.H. Shin, C.H. Cho, Production of hydrogen-rich synthetic gas from low-grade coals by microwave steam-plasmas, Int. J. Hydrogen Energy 39 (2014) 4351–4355. https://doi.org/https://doi.org/10.1016/j.ijhydene.2014.01.020.
[59] W.-C. Lin, C.-H. Tsai, D.-N. Zhang, S.-S. Syu, Y.-M. Kuo, Recycling of aluminum dross for producing calcinated alumina by microwave plasma, Sustain. Environ. Res. 32 (2022) 50. https://doi.org/10.1186/s42834-022-00160-9.
[60] G.S.J. Sturm, A.N. Muñoz, P. V Aravind, G.D. Stefanidis, Microwave-Driven Plasma Gasification for Biomass Waste Treatment at Miniature Scale, IEEE Trans. Plasma Sci. 44 (2016) 670–678. https://doi.org/10.1109/TPS.2016.2533363.
[61] P. Khongkrapan, P. Thanompongchart, N. Tippayawong, T. Kiatsiriroat, Microwave plasma assisted pyrolysis of refuse derived fuels, Cent. Eur. J. Eng. 4 (2014) 72–79. https://doi.org/10.2478/s13531-013-0142-5.
[62] S.J. Yoon, Y.M. Yun, M.W. Seo, Y.K. Kim, H.W. Ra, J.-G. Lee, Hydrogen and syngas production from glycerol through microwave plasma gasification, Int. J. Hydrogen Energy 38 (2013) 14559–14567. https://doi.org/https://doi.org/10.1016/j.ijhydene.2013.09.001.
[63] K.C. Lin, Y.-C. Lin, Y.-H. Hsiao, Microwave plasma studies of Spirulina algae pyrolysis with relevance to hydrogen production, Energy 64 (2014) 567–574. https://doi.org/https://doi.org/10.1016/j.energy.2013.09.055.
[64] M. Hu, W. Deng, Y. Su, L. Wang, G. Chen, Production of hydrogen-rich syngas through microwave-assisted gasification of sewage sludge in steam-CO2 atmosphere, Fuel 357 (2024) 129855. https://doi.org/https://doi.org/10.1016/j.fuel.2023.129855.
[65] B. Ibrahimoglu, M.Z. Yilmazoglu, Numerical modeling of a downdraft plasma coal gasifier with plasma reactions, Int. J. Hydrogen Energy 45 (2020) 3532–3548. https://doi.org/https://doi.org/10.1016/j.ijhydene.2018.12.198.
[66] E. Delikonstantis, G. Sturm, A.I. Stankiewicz, A. Bosmans, M. Scapinello, C. Dreiser, O. Lade, S. Brand, G.D. Stefanidis, Biomass gasification in microwave plasma: An experimental feasibility study with a side stream from a fermentation reactor, Chem. Eng. Process. - Process Intensif. 141 (2019) 107538. https://doi.org/https://doi.org/10.1016/j.cep.2019.107538.
[67] Y.C. Hong, S.J. Lee, D.H. Shin, Y.J. Kim, B.J. Lee, S.Y. Cho, H.S. Chang, Syngas production from gasification of brown coal in a microwave torch plasma, Energy 47 (2012) 36–40. https://doi.org/https://doi.org/10.1016/j.energy.2012.05.008.
[68] Y. Tanaka, Y. Yokomizu, M. Ishikawa, T. Matsumura, Particle composition of high-pressure SF/sub 6/ plasma with electron temperature greater than gas temperature, IEEE Trans. Plasma Sci. 25 (1997) 991–995. https://doi.org/10.1109/27.649615.

---

*Corresponding author*

*e-mail:* satyanada@dese.iitd.ac.in (Satyananda Kar)